\documentclass[aip,jcp,reprint,letterpaper,twocolumn]{revtex4-1}
\usepackage{amssymb}
\usepackage{amsmath}
\usepackage[usenames,dvipsnames]{xcolor}
\usepackage{graphicx}
\usepackage{float}
\usepackage[normalem]{ulem}
\usepackage[top=1 in, bottom=1 in]{geometry}
\usepackage{natbib}

\begin{document}

\title
{xDH double hybrid functionals can be qualitatively incorrect for non-equilibrium geometries: Dipole moment inversion and barriers to radical-radical association using XYG3 and XYGJ-OS}
\author{Diptarka Hait}
\affiliation
{{Kenneth S. Pitzer Center for Theoretical Chemistry, Department of Chemistry, University of California, Berkeley, California 94720, USA}}

\author{Martin Head-Gordon}
\email{mhg@cchem.berkeley.edu}
\affiliation
{{Kenneth S. Pitzer Center for Theoretical Chemistry, Department of Chemistry, University of California, Berkeley, California 94720, USA}}
\affiliation{Chemical Sciences Division, Lawrence Berkeley National Laboratory, Berkeley, California 94720, USA}
	
\begin{abstract}
		Double hybrid (DH) density functionals are amongst the most accurate density functional approximations developed so far, largely due to incorporation of correlation effects from unoccupied orbitals via second order perturbation theory (PT2). The xDH family of DH functionals calculate energy directly from orbitals optimized by a lower level approach like B3LYP, without self-consistent optimization. XYG3 and XYGJ-OS are two widely used xDH functionals that are known to be quite accurate at equilibrium geometries. Here, we show that the XYG3 and XYGJ-OS functionals can be ill behaved for stretched bonds well beyond the Coulson-Fischer point, predicting unphysical dipole moments and humps in potential energy curves for some simple systems like the HF molecule. Numerical experiments and analysis show these failures are not due to PT2. Instead, a large mismatch at stretched bond-lengths between the reference B3LYP orbitals and the optimized orbitals associated with the non-PT2 part of XYG3 lead to an unphysically large non-Hellman-Feynman contribution to first order properties like forces and electron densities.  
\end{abstract}
	\maketitle

Density functional theory is widely used for electronic structure calculations as it tends to yield sufficiently accurate results for a significantly lower computational cost relative to correlated wave function approaches\cite{mardirossian2017thirty,jones2015density}. However, approximate local functionals are often known to fail in describing systems with charge delocalization\cite{Ruzsinszky2006,Dutoi2006,cohen2011challenges} on account of missing important nonlocal information. Orbital dependent nonlocal exchange and correlation components are thus often hybridized with local exchange and correlation to produce more sophisticated functionals in the hope of attaining higher accuracy. Such functionals are typically classified as hybrids if they only contain nonlocal Hartree-Fock like exchange  and as double hybrids if both nonlocal orbital dependent exchange and correlation (typically from second order perturbation theory or RPA) are included. Double hybrid (DH) functionals consequently occupy the fifth rung of Jacob's ladder\cite{perdew2001jacob} and are amongst the most accurate density functionals developed so far, yielding highly accurate predictions of energies\cite{goerigk2017look}, dipole moments\cite{hait2017accurate} and other properties\cite{su2014fractional,su2016beyond}. This makes DH functionals very attractive for electronic structure calculations, despite a somewhat higher computational complexity due to the dependence on unoccupied orbitals for nonlocal correlation. 

The energy of a DH functional constructed from a hybrid part, $E_{{\rm{hyb}}}$ and the scaled second order perturbation theory (PT2) correlation energy, $E_{{\rm{PT2}}}$, may be written in two equivalent ways:
\begin{equation}
{E_{{\rm{DH}}}} = {E_{{\rm{hyb}}}} + {E_{{\rm{PT2}}}} = E_{{\rm{DH}}}^{{\rm{OO}}} + \Delta {E_{{\rm{DH}}}}
\label{DH}
\end{equation} 
The first generic form does not specify how the orbitals are optimized, while the second identifies the energy minimized via orbital optimization (OO) as $E_{{\rm{DH}}}^{{\rm{OO}}}$ plus a non-SCF correction, $\Delta {E_{{\rm{DH}}}}$. The first form is standard, while the second form is appropriate for considering the evaluation of molecular properties, as will be our purpose later.

With regard to choice of orbitals, three distinct OO approaches have been developed so far, which we will label as tDH, xDH and OO-DH. The truncated DH (tDH) examples like B2PLYP\cite{grimme2006semiempirical} optimize orbitals with a truncated Kohn-Sham (KS)\cite{kohn1965self} functional that is simply $E_{{\rm{tDH}}}^{{\rm{OO}}} = {E_{{\rm{hyb}}}}$, so that the non-SCF correction is {solely} the PT2 term {(i.e.} $\Delta {E_{{\rm{tDH}}}} = {E_{{\rm{PT2}}}}${)}. A later approach {proposed by} Zhang, Xu and Goddard\cite{zhang2009doubly} (xDH) uses orbitals optimized by a reference (ref) lower rung functional, like B3LYP\cite{b3lyp}, so that $E_{{\rm{xDH}}}^{{\rm{OO}}} = {E_{{\rm{ref}}}}$ and the non-SCF correction becomes $\Delta {E_{{\rm{xDH}}}} = \left( {{E_{{\rm{hyb}}}} - {E_{{\rm{ref}}}}} \right) + {E_{{\rm{PT2}}}}$. XYG3\cite{zhang2009doubly} and XYGJ-OS\cite{zhang2011fast} are two well known examples of this category. Finally, it is also possible\cite{peverati2013orbital} to use the total DH energy for orbital optimization (OO-DH) in a manner similar to orbital optimized MP2\cite{lochan2007orbital,neese2009assessment}, so that $\Delta {E_{{\rm{OO-DH}}}}=0$. OO-DH functionals are not yet widely used.  

Both the tDH and xDH approaches have certain drawbacks. The truncated KS functional used for orbital optimization in tDH, $E_{{\rm{tDH}}}^{{\rm{OO}}} = {E_{{\rm{hyb}}}}$, does not have a complete description of correlation, which will cause effects that include early onset of spin polarization. On the other hand, the xDH approach relies on the adiabatic connection\cite{zhang2009doubly}, which may not hold if the lower rung orbital generating functional like B3LYP yields insufficiently accurate densities, as has been suggested by a number of recent studies\cite{medvedev2017density,brorsen2017accuracy,hait2017accurate}. Nonetheless, both approaches have yielded functionals highly accurate for equilibrium properties\cite{goerigk2017look,zhang2009doubly,zhang2011fast,hait2017accurate} like B2GPPLYP\cite{karton2008highly} (tDH) and XYG3 (xDH).

Our recent study on dipole moments\cite{hait2017accurate}, however, suggested that the picture might not be so rosy for xDH functionals away from equilibrium. Both XYG3 and XYGJ-OS yielded dipoles on the order of $1-2$ D for the HF molecule at internuclear separations of $2.75-4.25$ {\AA}, versus a CCSD(2)\cite{Gwaltney2000,gwaltney2001second} reference value of $0.1$ D or less. The polarity furthermore was reversed from H$^+$F$^-$ to the unphysical H$^-$F$^+$. Similar unphysical behavior was seen for stretched HCl, ClF, LiH and CH$_3$F (along the C-F bond breaking coordinate), suggesting that such  {density} artifacts for stretched bonds is a rather general issue with these xDH functionals. Practical implications {aside}, this also raises several other questions that we address here: (i) Are only  {densities} affected or can other xDH properties also behave poorly at stretched geometries? (ii) What is the origin of this peculiar  behavior? (iii) Are tDH  functionals affected {as well}? 

\begin{figure}[htb!]
	\vspace*{-10pt}
	\includegraphics[width=\linewidth]{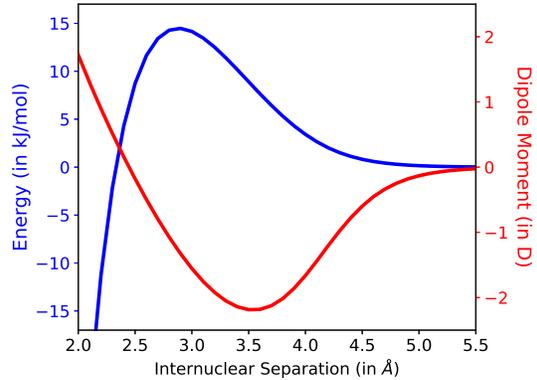}
	\vspace*{-10pt}
	\caption{Unphysical local extrema in the XYG3 dissociation curve of HF. The energy is relative to the dissociation limit and a polarity of H$^+$F$^-$ gives a positive dipole.}
	\vspace*{-5pt}
	\label{fig:hf-xyg3both}
\end{figure}

An examination of potential energy surfaces generated by XYG3 and XYGJ-OS for these species revealed that several have an unphysical local maximum at stretched geometries, resulting in a spurious \textit{barrier} for homolytic bond formation. This is illustrated in Fig \ref{fig:hf-xyg3both} for the HF molecule. Similar features were seen even for some non-polar species like F$_2$. The energy barrier is not very large ($\approx 3-15$ kJ/mol) relative to the bond strength but is chemically significant. Furthermore, the local energy maxima and the unphysical dipole moments occur in roughly the same region of the dissociation curve (as seen in Fig \ref{fig:hf-xyg3both}) suggesting that they may have similar origins. tDH functionals employing self-consistently optimized truncated KS orbitals like B2GPPLYP do not appear to have such features, {indicating} that the origin may lie in the xDH recipe itself. 

\onecolumngrid

\begin{figure*}[bht!]
	\vspace*{-5pt}
	\begin{minipage}{0.49\textwidth}
		\includegraphics[width=\linewidth]{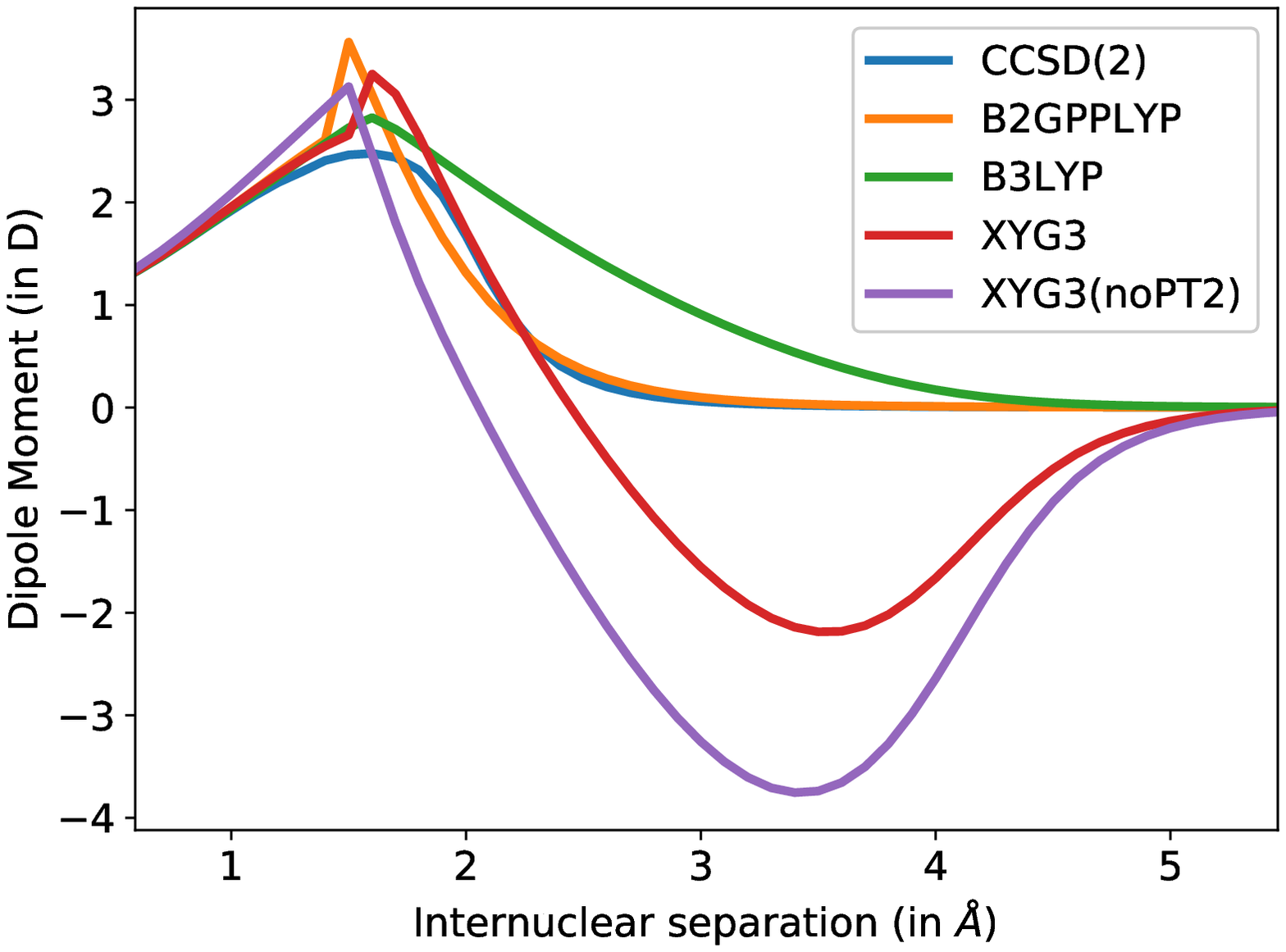}
	\end{minipage}
	\begin{minipage}{0.49\textwidth}
		\includegraphics[width=\linewidth]{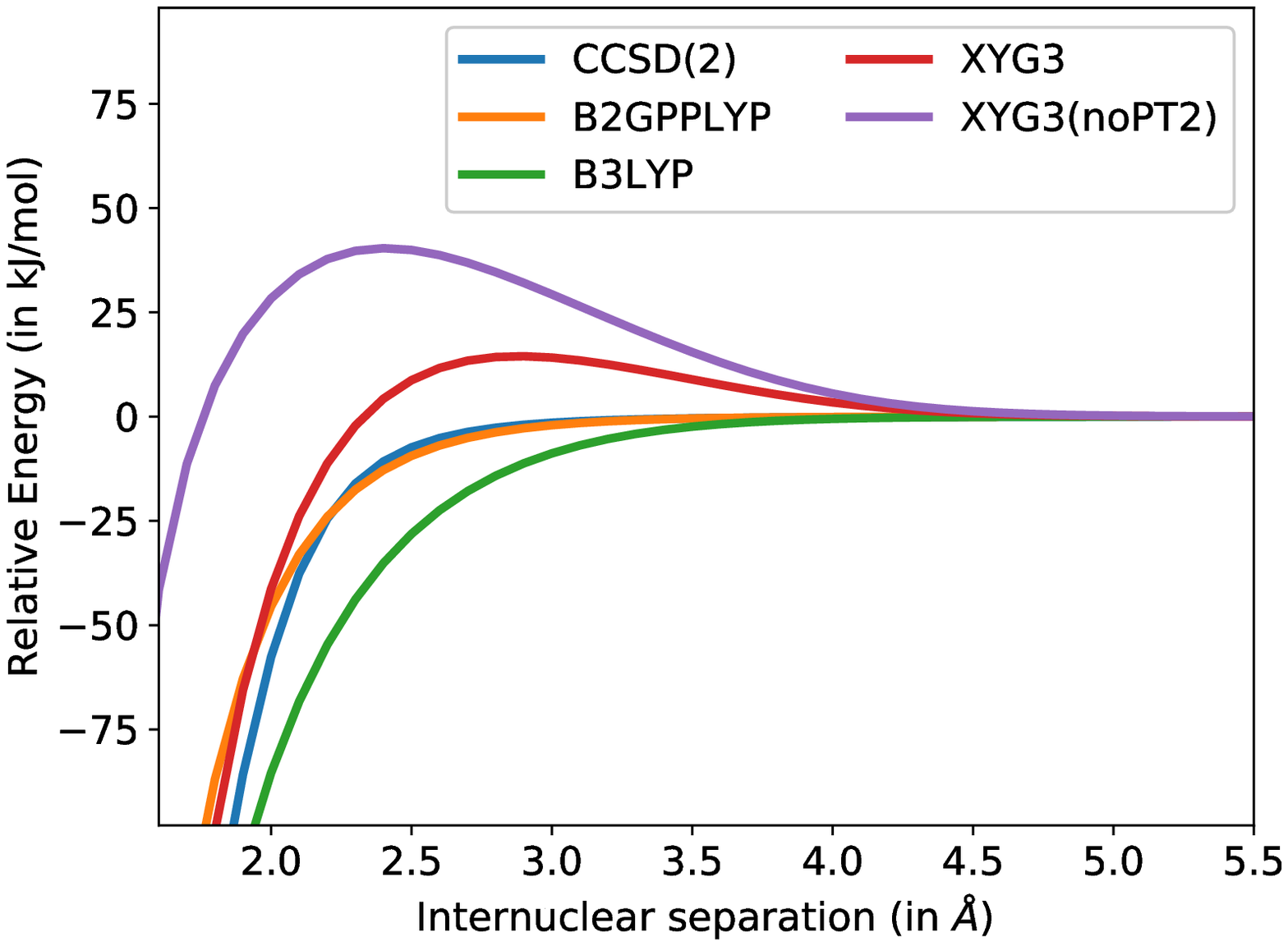}
	\end{minipage}
	\vspace*{-10pt}
	\caption{Dipole moments and potential energies predicted by XYG3 with and without PT2, along with the reference CCSD(2) , B3LYP and B2GPPLYP values. A polarity of H$^+$F$^-$ gives a positive dipole, and all the energy curves are referenced to the dissociation limit. The XYG3 artifacts are enlarged by removing the PT2 term, and the B3LYP curves approach the atomic asymptotes very slowly compared to CCSD(2).}
	\vspace*{-10pt}
	\label{fig:hf-xyg3dip}
\end{figure*}
\twocolumngrid


{We investigate the origins of this unphysical behavior through a detailed study on the HF dissociation curve. } Fig \ref{fig:hf-xyg3dip} plots  XYG3 {dipoles and potential energies} {for this system, along with} results from the reference (B3LYP), a tDH functional (B2GPPLYP), and a reliable wavefunction method (CCSD(2)). While not shown, XYGJ-OS exhibits similar behavior. The dipole curves  {predicted by} the two DH functionals {have} discontinuities for small bond-stretches, which arise from known N-representability violations in unrestricted MP2 and DH functionals\cite{Kurlancheek2009,hait2017accurate} around the Coulson-Fischer (CF)\cite{coulson1949xxxiv} point of the SCF method. The CF point lies between between $1.35$ {\AA} (Hartree-Fock) and $1.6$ {\AA} (B3LYP) for HF, where the sign of the dipole moment is still correct. The unphysical local extrema of XYG3 are found beyond $2$ {\AA} internuclear separation, indicating that the dipole inversion (and the energy maximum) is not directly connected to {this} N-representability issue, but arises from something else. The innocence of {$E_{{\rm{PT2}}}$} itself is made evident by the observation from Fig \ref{fig:hf-xyg3dip} that the unphysical local extremum is more pronounced when {$E_{{\rm{PT2}}}$} is deleted.  {$E_{{\rm{PT2}}}$} was therefore partially correcting these issues in the full functional. 



\begin{figure}[h!]
	\vspace*{-10pt}
	\includegraphics[width=\linewidth]{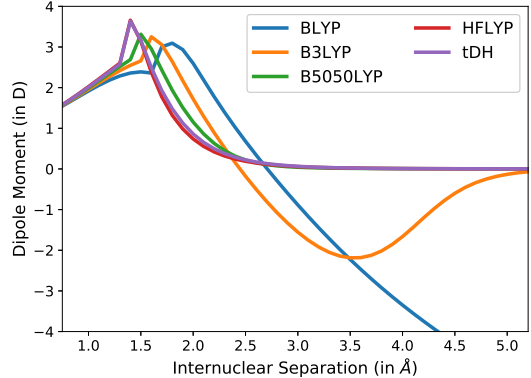}
	\vspace*{-10pt}
	\caption{The effect of different choices of reference orbitals on dipole moment curves evaluated with the XYG3 double hybrid energy functional. Using BLYP orbitals dramatically exaggerates the dipole inversion (the dissociation limit has a residual partial negative charge on H and a corresponding positive charge on F) relative to using B3LYP orbitals (as XYG3 is defined); while orbitals from B5050LYP, HFLYP or the non-PT2 part of XYG3 remove the dipole inversion.}
	\vspace*{-10pt}
	\label{fig:ref_orbitals}
\end{figure}

There are two possible sources for this unphysical behavior of the non-PT2 portion, $E_{{\rm{hyb}}}$, of the xDH functionals: either $E_{{\rm{hyb}}}$ is intrinsically unphysical or there is a major mismatch between the reference B3LYP orbitals and those which minimize $E_{{\rm{hyb}}}$. The former can be ruled out because turning XYG3 into a tDH functional (i.e. replacing B3LYP orbitals by $E_{{\rm{hyb}}}$ optimized orbitals) yields energies and dipoles that are qualitatively acceptable, as shown in Fig. \ref{fig:ref_orbitals}. Lack of self-consistency between B3LYP orbitals and $E_{{\rm{hyb}}}$ is thus the origin of these unphysical local extrema in the dissociation curves. A strong dependence of the dipole curves on choice of reference orbitals is established in Fig. \ref{fig:ref_orbitals}, where BLYP\cite{b88,lyp} ($0\%$ exact exchange) orbitals yield far worse behavior than XYG3's default use of B3LYP ($20\%$ exact exchange) orbitals . On the other hand, using orbitals with higher exact exchange content (i.e. closer to XYG3's $E_{{\rm{hyb}}}$ with $\approx80\%$ exact exchange) such as B5050LYP\cite{b5050lyp} ($50\%$ exact exchange) and HFLYP\cite{lyp} ($100\%$ exact exchange)  entirely removes the issue.

Having {numerically} established the origin of the density and property artifact, let us next analyze it more closely. Returning to the second form of \eqref{DH}, a first order property, $E_{{\rm{DH}}}^x$ like the dipole moment or force includes direct (Hellman-Feynman) derivative contributions that we can denote as $E_{{\rm{DH}}}^{\left( x \right)}$, and non-Hellman-Feynman terms that reflect how the orbital degrees of freedom, $\theta$, change as a function of the perturbation, $x$. These terms arise only from the non-SCF part, $\Delta {E_{{\rm{DH}}}}$, of $E_{{\rm{DH}}}$. In detail, the first order derivative (property) is given by:
\begin{equation}
E_{{\rm{DH}}}^x = E_{{\rm{DH}}}^{\left( x \right)} + \Delta E_{{\rm{DH}}}^\theta {\theta ^x} = E_{{\rm{DH}}}^{\left( x \right)} + z \cdot {\left( {E_{{\rm{DH}}}^{{\rm{OO}}}} \right)^{\theta x}}
\label{eq:1st_order_property}
\end{equation}
In the second form, $z$ is the so-called z-vector\cite{handy1984evaluation} which corresponds to the occupied-virtual or response block of the DH density matrix, and is the solution to the response equation in terms of the orbital-optimized energy: ${\left( {E_{{\rm{DH}}}^{{\rm{OO}}}} \right)^{\theta \theta '}}z + \Delta E_{{\rm{DH}}}^\theta  = 0$ {(where ${\left({E_{{\rm{DH}}}^{{\rm{OO}}}} \right)^{\theta \theta '}}$ is the Hessian of the orbital optimized energy $E_{{\rm{DH}}}^{{\rm{OO}}}$, with respect to orbital rotation and $\Delta E_{{\rm{DH}}}^\theta $ is the rate of change of the non orbital optimized energy with orbital rotation).}

The response density, $z$, is strictly zero for OO-DH functionals (one of their main benefits). It is evidently small for the tDH functionals tested here, but large for the xDH functionals, XYG3 and XYGJ-OS, that use B3LYP reference orbitals. The B3LYP orbitals give qualitatively correct behavior on their own (though Fig \ref{fig:hf-xyg3dip} suggests that they exhibit significant density delocalization and energy lowering at stretched bond lengths relative to the CCSD(2) reference values). The critical thing, however, is that they are not optimized for the non-PT2 truncated KS part of xDH functionals, leading to a non Hellman-Feynman contribution (second term of \eqref{eq:1st_order_property}) to both dipoles and forces from the non-PT2 part as well. This term can be large if there is a major mismatch between the orbitals and the non-PT2 part of the functional. This is completely consistent with the numerically identified origin of the unphysical behavior, as presented in Fig. \ref{fig:ref_orbitals}.

$z =-\left[{\left( {E_{{\rm{DH}}}^{{\rm{OO}}}} \right)^{\theta \theta '}}\right]^{-1} \Delta E_{{\rm{DH}}}^\theta $ indicates that $z$ might be unphysically large either due to a poorly conditioned $\left( {E_{{\rm{DH}}}^{{\rm{OO}}}} \right)^{\theta \theta '}$ or a large $\Delta E_{{\rm{DH}}}^\theta$. The smallest eigenvalues of ${\left({E_{{\rm{DH}}}^{{\rm{OO}}}} \right)^{\theta \theta '}}$ however do not appear to be particularly different between methods in Fig \ref{fig:ref_orbitals}, indicating that the unphysicalities in $z$ stem at least in part from $\Delta E_{{\rm{DH}}}^\theta$, enabling a chemical interpretation of the mathematics.  At the very stretched geometries where XYG3 shows problems, the B3LYP orbitals are too delocalizing relative to the truncated hybrid functional (which has $\approx 80\%$ exact exchange vs $20\%$ in B3LYP). The orbital gradient of the XC difference,  $\Delta E_{{\rm{DH}}}^\theta$ therefore corrects the reference orbitals (and density) to be more localizing. Self-consistent iterations that treat XYG3 as an OO-DH rigorously zero this quantity. $\Delta E_{{\rm{DH}}}^\theta$ is also made smaller by choices of reference orbitals that are closer to OO-DH such as {by} treating XYG3 as a tDH functional. However, in XYG3 used as an xDH with B3LYP orbitals, the critical problem is that this linear response term leads to a substantial \emph{overcorrection} in the density for $R > 2.5 ${\AA}, which in turn leads to the density and force artifacts already discussed. Overall, the artifacts identified in this work originate from a breakdown in the linear response approximation to OO-DH under conditions such as highly stretched bonds, where there can be a substantial orbital mismatch between the reference orbitals and the optimal ones. Asymptotically incorrect reference densities could in particular lead to a wrong dissociation limit, as is observed for the case of C-F bond breaking in CH$_3$F, where partial charges persist even at large separations for XYG3 applied on a B3LYP reference (which itself has residual charge of the opposite polarity). A large non Hellman-Feynman term may even cause the one-particle density  matrix to become non N-representable in extreme cases.

While the breakdown of XYG3 and XYGJ-OS is interesting, and potentially valuable as a guide to developing  DH functionals in the future, it is important to clarify two points. First, XYG3 and XYGJ-OS are generally very good functionals for the study of chemistry around equilibrium molecular configurations, as is consistent with existing studies\cite{su2014fractional,su2016beyond,zhang2009doubly,zhang2011fast}.  That is consistent with the way in which the few parameters in these xDH functionals were trained, and is consistent with the expectation that there will be only a small orbital mismatch under such conditions. Thus our present results therefore are not in themselves a reason to abandon routine use of XYG3 and XYGJ-OS in the equilibrium regime, or, when there is no reason to expect a significant orbital mismatch. Our results are certainly a reason to recommend that future attempts to develop widely applicable DH type functionals (especially xDH) should carefully take into account behavior at non-equilibrium configurations.

A second point to clarify is that our results do not imply that unphysical behavior will always occur at non-equilibrium configurations as it is quite possible to have a small non Hellman-Feynman term. Indeed, dissociation curves for H$_2$ and Li$_2$ appear to be perfectly well described by both XYG3 and XYGJ-OS. Furthermore, the PT2 contribution also appears to act in the opposite direction from the non Hellman-Feynman KS term, and may in fact be successful in ameliorating unphysical behavior in some cases. It is beyond the scope of this letter to develop a quantitative metric that can be employed to determine a priori whether XYG functionals (or indeed any xDH functional) will give physical results or not. However, it is clear from our present results that there is reason to be cautious wherever a large difference between reference densities and xDH truncated KS self-consistent densities might be suspected.

All calculations performed with a development version of the Q-Chem 4.0 package\cite{QCHEM4}, and unrestricted orbitals were employed in each case. Density functional calculations were done with the aug-cc-pCVQZ\cite{dunning1989gaussian,woon1995gaussian,peterson2002accurate,prascher2011gaussian} basis (except for CH$_3$F, for which aug-cc-pCVTZ was used), and all electron CCSD(2) calculations were done with the aug-cc-pCVTZ basis. This research was supported by the Director, Office of Science, Office of Basic Energy Sciences, of the U.S. Department of Energy under Contract No. DE-AC02-05CH11231. D.H. was also funded by a Berkeley Fellowship. 
\bibliography{references}
\end{document}